%% file: main.tex
\documentclass[%
reprint,
jmp,
superscriptaddress,
nofootinbib,
amsmath,amssymb,
table
]{revtex4-2}
\usepackage[english]{babel}
\usepackage{graphicx}
\usepackage{dcolumn}
\usepackage{bm}
\usepackage[version=4]{mhchem}
\usepackage{svg}
\usepackage{multirow}
\usepackage{comment}
\usepackage{tabularx}
\usepackage{array}
\usepackage{makecell}
\usepackage{amssymb}
\usepackage{float}
\usepackage{afterpage}
\usepackage[utf8]{inputenc}
\usepackage{braket}
\usepackage{amsmath}
\usepackage{amsfonts}
\usepackage{booktabs} 
\usepackage{titlesec}
\usepackage{dsfont}
\usepackage{pifont}
\usepackage{adjustbox}
\usepackage[normalem]{ulem}
\usepackage{tabularray}
\usepackage[left=0.8in,right=0.8in,top=0.8in,bottom=0.8in]{geometry}
\usepackage{upgreek}
\usepackage{textgreek}
\usepackage[super]{nth}
\usepackage{csquotes}
\usepackage{xcolor}
\usepackage{xr}
\usepackage{appendix}
\usepackage{listings}
\usepackage{hhline}
\definecolor{codegreen}{rgb}{0,0.6,0}
\definecolor{codegray}{rgb}{0.5,0.5,0.5}
\definecolor{codepurple}{rgb}{0.58,0,0.82}
\definecolor{backcolour}{rgb}{0.95,0.95,0.92}

\lstdefinestyle{bashstyle}{
    backgroundcolor=\color{backcolour},   
    commentstyle=\color{codegreen},
    keywordstyle=\color{magenta},
    numberstyle=\tiny\color{codegray},
    stringstyle=\color{codepurple},
    basicstyle=\ttfamily\footnotesize,
    breakatwhitespace=false,         
    breaklines=true,                 
    captionpos=b,                    
    keepspaces=true,                 
    numbers=left,                    
    numbersep=5pt,                  
    showspaces=false,                
    showstringspaces=false,
    showtabs=false,                  
    tabsize=2,
    frame=single,
    rulecolor=\color{black}
}

\newcolumntype{Y}{>{\centering\arraybackslash}X}

\renewcommand{\thefootnote}{\fnsymbol{footnote}}

\definecolor{myblue}{RGB}{71,120,207}
\definecolor{mygreen}{RGB}{106,204,100}
\definecolor{myred}{RGB}{213,95,95}

\date{\today}

\makeatletter
\def\@email#1#2{%
 \endgroup
 \patchcmd{\titleblock@produce}
  {\frontmatter@RRAPformat}
  {\frontmatter@RRAPformat{\produce@RRAP{*#1\href{mailto:#2}{#2}}}\frontmatter@RRAPformat}
  {}{}
}%
\makeatother

\begin{document}

\preprint{AIP/123-QED}

\title{Computing solvation free energies of small molecules with experimental accuracy}

\author{J. Harry Moore\textsuperscript{\dag}}
\affiliation{Engineering Laboratory, University of Cambridge, Cambridge, CB2 1PZ, UK}
\affiliation{Ångström AI, 2325 3rd Street, San Francisco, CA 94107, USA}
\author{Daniel J. Cole}
\affiliation{School of Natural and Environmental Sciences, Newcastle University, Newcastle upon Tyne NE1 7RU, UK}

\author{G\'abor Cs\'anyi}
\affiliation{Engineering Laboratory, University of Cambridge, Cambridge, CB2 1PZ, UK}
\affiliation{Ångström AI, 2325 3rd Street, San Francisco, CA 94107, USA}

\begin{abstract}
\section*{Abstract}

Free energies play a central role in characterising the behaviour of chemical systems and are among the most important quantities that can be calculated by molecular dynamics simulations. Solvation free energies in various organic solvents, in particular, are well-studied physicochemical properties of drug-like molecules and are commonly used to assess and optimise the accuracy of nonbonded parameters in empirical forcefields, and also as a fast-to-compute surrogate of performance for protein-ligand binding free energy estimation. Machine learned potentials (MLPs) show great promise as more accurate alternatives to empirical forcefields, but are not readily decomposed into physically motivated functional forms, which has thus far rendered them incompatible with standard alchemical free energy methods that manipulate individual pairwise interaction terms. 
However, since the accuracy of free energy calculations is highly sensitive to the forcefield, this is a key area in which MLPs have the potential to address the shortcomings of empirical forcefields. 
In this work, we introduce an efficient alchemical free energy protocol that enables calculations of rigorous free energy differences in condensed phase systems modelled entirely by MLPs. Using a pretrained, transferrable, alchemically equipped MLP model, we demonstrate sub-chemical accuracy for the solvation free energies of a wide range of organic molecules.
\end{abstract}

\maketitle
\def\thefootnote{\dag}\footnotetext{jhm72@cam.ac.uk}

\section{Introduction}
Free energies are arguably the most important quantities accessible by molecular simulation in soft matter and play an important role in computational drug discovery~\cite{courniaRelativeBindingFree2017, schindlerLargeScaleAssessmentBinding2020}.  Free energy methods rely on rigorous thermodynamic protocols, enabling prediction of protein-ligand binding~\cite{wangAccurateReliablePrediction2015, gapsysLargeScaleRelative2020a}, small molecule solvation and solubility~\cite{loefflerReproducibilityFreeEnergy2018}, organic crystal polymorphs ranking~\cite{abrahamStatisticalMechanicalApproximations2020} and protein residue mutation~\cite{gapsysAccurateRigorousPrediction2016}. 

Driven by the widespread availability of efficient GPU hardware, highly optimised molecular dynamics (MD) software~\cite{eastmanOpenMMMolecularDynamics2023} and carefully parameterised forcefields~\cite{boothroydDevelopmentBenchmarkingOpen2023a, wangAutomaticAtomType2006a}, free energy calculations have become an industry standard tool, accounting for a large fraction of the computation performed in pharmaceutical R\&D. Relative binding free energy (RBFE) calculations in particular have emerged as a crucial aspect of structure-based drug discovery, helping to screen and rank congeneric series of compounds during hit-to-lead and lead optimisation campaigns~\cite{courniaRelativeBindingFree2017}.

Whilst various methods for calculating free energies for small molecules via MD have been proposed, alchemical transformations have emerged as the \textit{de facto} standard approach, allowing free energy estimates to be obtained in a few GPU hours per compound~\cite{choderaAlchemicalFreeEnergy2011, meyBestPracticesAlchemical2020}.
Rather than explicitly simulating a compound undergoing a change of state as it would happen in reality (which would be computationally prohibitive), alchemical calculations take advantage of the fact that the free energy is a state function and so the free energy difference between two state points is independent of the pathway connecting them.  By introducing an alchemical parameter, $\lambda$,  a new Hamiltonian is constructed as a linear combination of the Hamiltonians describing the two end states,
\begin{equation}
    H(\Vec{r}, \lambda) = \lambda H_1(\Vec{r}) + (1 - \lambda)H_0(\Vec{r})
\end{equation}
The free energy difference associated with the transformation can then be computed by a variety of estimators, for example by thermodynamic integration:
\begin{equation}
    \Delta G = \int_0^1 \left \langle \frac{\partial H(\vec{r}, \lambda)}{\partial \lambda} \right \rangle_{\lambda} d\lambda
    \label{eq:thermoint}
\end{equation}
where $\langle \dots \rangle_{\lambda}$ denotes the ensemble average corresponding to the Hamiltonian $H(\lambda)$. Using such artificial pathways leads to enormous computational savings, since the sampling required to explore intermediate state points is orders of magnitude less than it would be for a real pathway describing the physical process. 
In practice, the transformation is accomplished by identifying the atoms that change during the transformation (i.e. those atoms whose chemical element changes, or that disappear entirely) and interpolating the corresponding forcefield parameters between the end states.  This requires the use of so-called dummy atoms to model particles that disappear during the transformation. These atoms maintain their bonded terms but have their interactions with the surroundings switched off. This allows for transformations like morphing a methyl group to a hydrogen atom or isolating a molecule from its environment, while minimally perturbing the phase space required to be explored by the system.

Early attempts to perform free energy calculations with MD faced energy conservation and convergence issues due to the divergence of the nonbonded energy (usually a combination of Lennard-Jones and Coulomb contributions) when partially decoupled atoms overlap~\cite{squireMonteCarloSimulation1969, mezeiFreeEnergySimulationsa1986}. Beutler \textit{et al} were the first to address this issue by incorporating softening parameters to scale nonbonded interactions as a function of 
the alchemical parameter $\lambda$, thus averting singularities in the potential energy as atoms come into contact~\cite{beutlerAvoidingSingularitiesNumerical1994a}:
%
\begin{multline}
    U(\lambda, r) = 4\epsilon\lambda^n \Bigg[\Bigg.\left(\alpha_{LJ}(1-\lambda)^m + \left(\frac{r}{\sigma}\right)^6\right)^{-2} - \\ 
    \left(\alpha_{LJ}(1-\lambda)^m + \left(\frac{r}{\sigma}\right)^6\right)^{-1}\Bigg.\Bigg]
    \label{eqn:Beutler_softcore}
\end{multline}
where $\epsilon$ and $\sigma$ are the Lennard-Jones well depth and radius, respectively, and $\alpha_{LJ}$, $m$ and $n$ are positive, tunable constants that control the smoothness and decay of the softcore function.  In the original implementation, $\alpha_{LJ}=0.5$,  $n=4$ and $m=2$, although $m=n=1$ has been shown to reduce the variance of the free energy estimate~\cite{beutlerAvoidingSingularitiesNumerical1994a, steinbrecherNonlinearScalingSchemes2007, shirtsSolvationFreeEnergies2005}.
Although Beutler-type soft core potentials are commonly used in various simulation packages, it is worth noting that several alternative approaches have been suggested to further improve numerical stability and address specific failure modes~\cite{gapsysNewSoftCorePotential2012, leeImprovedAlchemicalFree2020}, and alternative nonbonded functional forms with a more natural soft core have also been tried~\cite{hortonTransferableDoubleExponential2023}. 

While these soft-core potentials have been widely successful, a significant limitation of current state-of-the-art free energy calculations lies in the accuracy of the empirical forcefields. Specifically, the constraints imposed by the functional form, omission of energetic contributions and restricted atom-type parameters in small molecule forcefields have a significant impact on accuracy~\cite{courniaRelativeBindingFree2017}.  Particularly notable are the widespread use of fixed-charge electrostatic models that do not account for geometry-dependent polarization, and the limited accuracy of torsional barriers, which often require refitting to bespoke DFT calculations of each new compound to achieve reasonable accuracy~\cite{harderOPLS3ForceField2016, abelAdvancingDrugDiscovery2017,hortonOpenForceField2022}.

In recent years, transferable machine learned potentials (MLP)s have been introduced as a compelling alternative to empirical forcefields, demonstrating significant advances in materials modelling and biomolecular simulation~\cite{devereuxExtendingApplicabilityANI2020,kovacsMACEOFFShortRangeTransferable2025, batatiaFoundationModelAtomistic2024, anstineAIMNet2NeuralNetwork2024, dengCHGNetPretrainedUniversal2023, unkePhysNetNeuralNetwork2019, kabyldaMolecularSimulationsPretrained2025}.
Building upon Behler and Parinello's initial work~\cite{Behler2007GeneralizedSurfaces}, the field has expanded rapidly, with many architectures being proposed to accurately model the QM potential energy surface. A particularly successful innovation has been the specific encoding of permutational, rotatational and translational symmetries~\cite{batatiaDesignSpaceEquivariant2022}.  These approaches have led to a series of data-efficient, universal potentials for biomolecular and materials simulation, enabling accurate prediction of atomistic and thermodynamic properties for a wide range of chemical compositions, albeit at an increased cost compared to empirical potentials.

Although MLPs can deliver significant accuracy improvements compared to empirical forcefields, their computational expense and the high sampling requirements of free energy calculations have meant that their application has been mostly restricted to corrective perturbations~\cite{coleMachineLearningBased2020,rufaChemicalAccuracyAlchemical2020, karwounopoulosInsightsChallengesCorrecting2024a}. In this approach, the intramolecular interactions of a subset of atoms (usually the organic molecule) are modelled by the MLP, while the remaining bonded interactions and all nonbonded interactions are modelled by the empirical forcefield. Although there is some evidence that this approach can modestly improve the accuracy of RBFE calculations, it does not address the parametrisation of nonbonded interactions, which are known to be crucial for obtaining accurate free energies~\cite{rufaChemicalAccuracyAlchemical2020}. Indeed, it has recently been shown that, for hydration free energies, MLP corrections with mechanical embedding fail to provide statistically significant accuracy improvements~\cite{karwounopoulosInsightsChallengesCorrecting2024a}.
It is worth noting that more sophisticated electrostatic embedding ML/MM schemes have recently been proposed~\cite{zinovjevElectrostaticEmbeddingMachine2023,morado2025, pultarNeuralNetworkPotential2025}. However, the performance of these methods in alchemical binding free energy calculations has not yet been established. 

In parallel, a handful of free energy methods that do not rely on alchemical transformations have been proposed, making the calculation conceptually more straightforward, easier to set up and more amenable to MLPs. One promising example uses the alchemical transfer method (ATM)~\cite{sabaneszariquieyEnhancingProteinLigand2024}, which relies on a coordinate transformation to interpolate between two physical end states~\cite{wuAlchemicalTransferApproach2021a, chenPerformanceAnalysisAlchemical2024}. It has been shown that, especially in combination with the ANI-2x potential for modelling the ligand intramolecular potential energy surface~\cite{devereuxExtendingApplicabilityANI2020}, this method is competitive with the commercially available FEP+ package~\cite{wangAccurateReliablePrediction2015}, which uses the empirical OPLS4 forcefield~\cite{luOPLS4ImprovingForce2021a}. This work still relies upon a mechanically embedded ML/MM Hamiltonian, and since both physical end states must be accommodated within the same simulation box, it is less efficient than alchemical perturbation methods due to the large amount of excess solvent required~\cite{sabaneszariquieyEnhancingProteinLigand2024}.

Given the potential for significant accuracy improvements, there is a need for rigorous and efficient alchemical free energy methods that are compatible with (i) describing the entire system using MLPs, and (ii)  existing well-tested free energy protocols and analysis packages (e.g. pymbar~\cite{shirtsStatisticallyOptimalAnalysis2008}, pmx~\cite{Seeliger2010pmx, Gapsys2015pmx}) that are currently in use with classical forcefields.  In this work, we present a pre-trained, transferable biomolecular MLP based on the recently introduced MACE-OFF potentials~\cite{kovacsMACEOFFShortRangeTransferable2025} that is uniquely equipped with scalable soft-core interactions, enabling numerically stable simulations of condensed phase systems with alchemically decoupled intermolecular interactions. We provide an efficient implementation of the method in the widely used OpenMM package~\cite{eastmanOpenMMMolecularDynamics2023}. Furthermore, we demonstrate the fast convergence and sub-chemical accuracy of our approach in the calculation of solvation free energies and distribution coefficients for a series of organic molecules. 

\section{Results \& Discussion}

\subsection{Alchemical simulations with MLPs}




To address the divergence of the potential energy predicted by a MLP fitted to DFT, as atoms overlap when sampling intermediate values of $\lambda$, we augmented the training set with dimer configurations containing artificially softened two-body interactions.  We further incorporated a $\lambda$ dependence via modification of the non-trainable parameters of the model architecture to scale nonbonded interactions between the solute and the solvent.  This combined approach results in $\lambda$-dependent many-body interactions that enable smooth and regularised interpolation between the coupled and decoupled end states. We note that, since only the nontrainable parameters are modulated during the simulation, no major changes to the training protocol are required. In all cases, the potential was trained corresponding to the $\lambda=1$ or fully interacting state, whilst the message scaling is only applied at inference time to the fitted potential.


\subsubsection{Softcore dimer curves}

\begin{figure}[H]
    \centering
    \includegraphics[width=0.8\columnwidth]{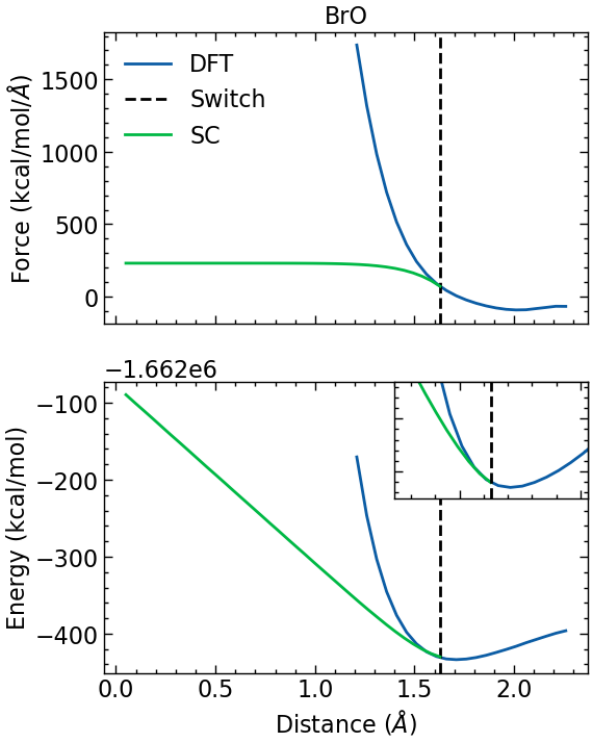}
    \caption{Construction of softcore dimer curve for the Br--O diatomic pair by matching the gradient at a set switching point to that given by DFT. }
    \label{fig:dft_softcore_curve}
\end{figure}

To construct the softcore dimer curves, we first computed dimer curves with DFT at the same $\omega$B97M-D3BJ/def2-TZVPPD level of theory as the SPICE dataset~\cite{eastman2023spice} for all combinations of the 10 elements covered by MACE-OFF~\cite{kovacsMACEOFFShortRangeTransferable2025}. For each pair, we started by matching the value and gradient of the DFT force with a polynomial of the form $ax^{10} + b$ that approaches a constant value as $r \to 0$ (Figure \ref{fig:dft_softcore_curve}). The softcore energies were then obtained by analytical integration of the polynomial to ensure consistent energy and force data. Without appropriate precautions, this approach can lead to artificial softening of the minimum of the dimer curve if the switching point is too close to the minimum. Conversely, a switching point that is too far from equilibrium would lead to large energy and force values for small separations, which are more challenging for the model to learn.

To balance these competing factors, the transition point was adjusted for each dimer curve to minimise the gradient of the force at the switching point, down to a minimum value of 3 eV/\AA{}\textsuperscript{2}, which in turn limited the maximum energy of pairwise interactions to a numerically stable value, typically around 300 kcal/mol above the minimum. However, by setting a minimum gradient for the switching point, the transition point was chosen so as not to alter the position and curvature of the energy minimum, thereby preserving the equilibrium properties of the potential. This therefore ensures that the softcore region is only explored during alchemical simulations in which the nonbonded interactions are (partially) decoupled, since this region of the energy landscape lies several hundred kcal/mol higher in energy than equilibrium. All switching point distances are tabulated in Section~\ref{SI:switching_points}.



The steeper curvature of tenth order functions enables smaller maximum forces and energies compared to lower-order polynomials that have been employed in previous softcore formulations~\cite{gapsysNewSoftCorePotential2012}. Using the \texttt{MACE-OFF24} training dataset~\cite{kovacsMACEOFFShortRangeTransferable2025} augmented with the synthetic dimer curves, we then fitted the softcore potential from scratch. Since it is expected that the learned radial embeddings will differ substantially from the MACE-OFF models due to the new softcore data, we found that this resulted in lower force and energy errors compared to transfer learning from the original \texttt{MACE-OFF} checkpoints.


The test set errors for the \texttt{MACE-OFF} data set were comparable to the performance of the original potentials~\cite{kovacsMACEOFFShortRangeTransferable2025}, indicating that the model's ability to learn energies and forces of equilibrium configurations is not affected by additional soft-core data (Section~\ref{sec:test_errors}).
We further confirmed that the predicted density of liquid water is unaffected by the new potential, with an average value of  $1.02\pm0.02$~g/mL at 298~K (compared to 0.98~g/mL for \texttt{MACE-OFF}24(M)).

It is worth noting that, whilst the introduction of softcore interactions is not expected to affect the ensemble sampled by the model at room temperature, since the curvature of the minimum is not affected and a substantial repulsive component is still learned by the model, this may not necessarily translate to high temperature regimes. In this case, the softcore region of the potential may be thermally accessible in the fully interacting state, rather than only being accessible in the partially decoupled states.

\subsubsection{Alchemical modification to MACE}
Whilst in principle the softcore formulation described above is sufficient to perform stable alchemical free energy calculations, in practice, the resulting linear formulation, where the functional form of the potential has no explicit dependence on $\lambda$, can lead to significant challenges with phase space overlap~\cite{buelensLinearscalingSoftcoreScheme2012}. 
Functions lacking this $\lambda$ dependence typically require complex reweighting strategies to adjust window spacing on the fly to maintain reasonable overlap between neighbouring replicas. To avoid these issues, we also modified the nonlearnable parameters in MACE to enable alchemical scaling of selected nonbonded interactions, resulting in a formulation analogous to classical soft-core forcefields ~\cite{beutlerAvoidingSingularitiesNumerical1994a, gapsysNewSoftCorePotential2012, leeImprovedAlchemicalFree2020}. By modifying only the nontrainable parameters of MACE, our approach enables alchemical simulations of arbitrary systems without additional fine tuning of the model or modification of the training protocol.

Whilst we introduce alchemical modifications in terms of the MACE architecture, we stress that our approach can be applied to any architecture that parameterises the local energy as a function of two-body terms, as has been recently demonstrated elsewhere~\cite{pleFoundationModelAccurate2025a}.
%
%
%
Here we describe the changes to the MACE architecture required to implement scaled nonbonded interactions. For a full description of the architecture, we direct the reader to the original publications~\cite{Batatia2022mace, batatiaDesignSpaceEquivariant2022}.

MACE predicts the total energy of a system as a sum of atom-wise contributions.  Each atom's environment is constructed from the relative coordinates and atomic numbers of its neighbours, up to a fixed cutoff.  These high body-order atomic features are efficiently constructed by taking tensor products of two-body terms that describe pairwise interactions.

For each atom, the intial features of its neighbours are combined with the interatomic displacement vectors, $\bm r_{ij}$, to form the one-particle basis $\phi_{ij,k \eta_{1} l_{3}m_{3}}^{(t)}$. The radial distance is used as an input into a learnable radial function $R(r_{ij})$ with several outputs that correspond to the different ways in which the displacement vector and the node features can be combined while preserving equivariance~\cite{wigner2012group}:
\begin{align}
  \label{eq:phi-basis-t}
  \phi_{ij,k \eta_{1} l_{3}m_{3}}^{(t)}(\lambda) &= 
    \sum_{l_1l_2m_1m_2} C_{\eta_1,l_1m_1l_2m_2}^{l_3m_3} \\ 
    \notag &\alpha_{ij} R_{k \eta_{1} l_{1}l_{2} l_{3}}^{(t)}(r_{ij}) \times 
       Y^{m_{1}}_{l_{1}} (\boldsymbol{\hat{r}}_{ij}) h^{(t)}_{j,kl_2m_2} 
\end{align}
where $Y_l^m$ are the spherical harmonics, and $C_{\eta_1,l_1m_1l_2m_2}^{l_3m_3}$ denotes the Clebsch-Gordan coefficients.
There are multiple ways of constructing an equivariant combination with a given symmetry $(l_3,m_3)$, and these multiplicities are enumerated by the path index $\eta_1$~\cite{dusson2022atomic, Batatia2022mace}.  

In a similar approach to other recently published work~\cite{namInterpolationDifferentiationAlchemical2024}, our formulation modifies MACE by including an additional factor of $\alpha_{ij}$ that scales selected two-body terms in the one-particle basis.  These terms are selected as those that cross the alchemical boundary, for example edges connecting atoms in the solute and solvent in the case of a solvation free energy calculation.

\begin{equation}\label{eq:alchemical_selector}
    \alpha_{ij} = 
    \begin{cases}  
      \lambda  &\text{if } i \in \text{solute} \oplus j \in \text{solute} \\
      1        & \text{else}  
    \end{cases}
\end{equation}

The one-particle basis functions $\phi_{ij,k \eta_{1} l_{3}m_{3}}^{(t)}$ are then passed through multiple layers of message passing, during which tensor products are taken to construct the many-body symmetric features.  
\begin{equation}
\label{eq:symmbasis_L1}
  {\bm B}^{(t),\nu}_{i,\eta_{\nu} k LM}
  = \sum_{{\bm l}{\bm m}} \mathcal{C}^{LM}_{\eta_{\nu} \bm l \bm m} \prod_{\xi = 1}^{\nu} A_{i,k l_\xi  m_\xi}^{(t)} 
\end{equation}.

This approach (named \texttt{MACE-OFF24-SC}) enables smoothly scaled many-body nonbonded interactions whilst maintaining the correct physical description of the end states, where the decoupled system's graph is identical to that of the separated components, ensuring consistent free energy calculations.

\begin{figure}
    \centering
    \includegraphics[width=0.8\columnwidth]{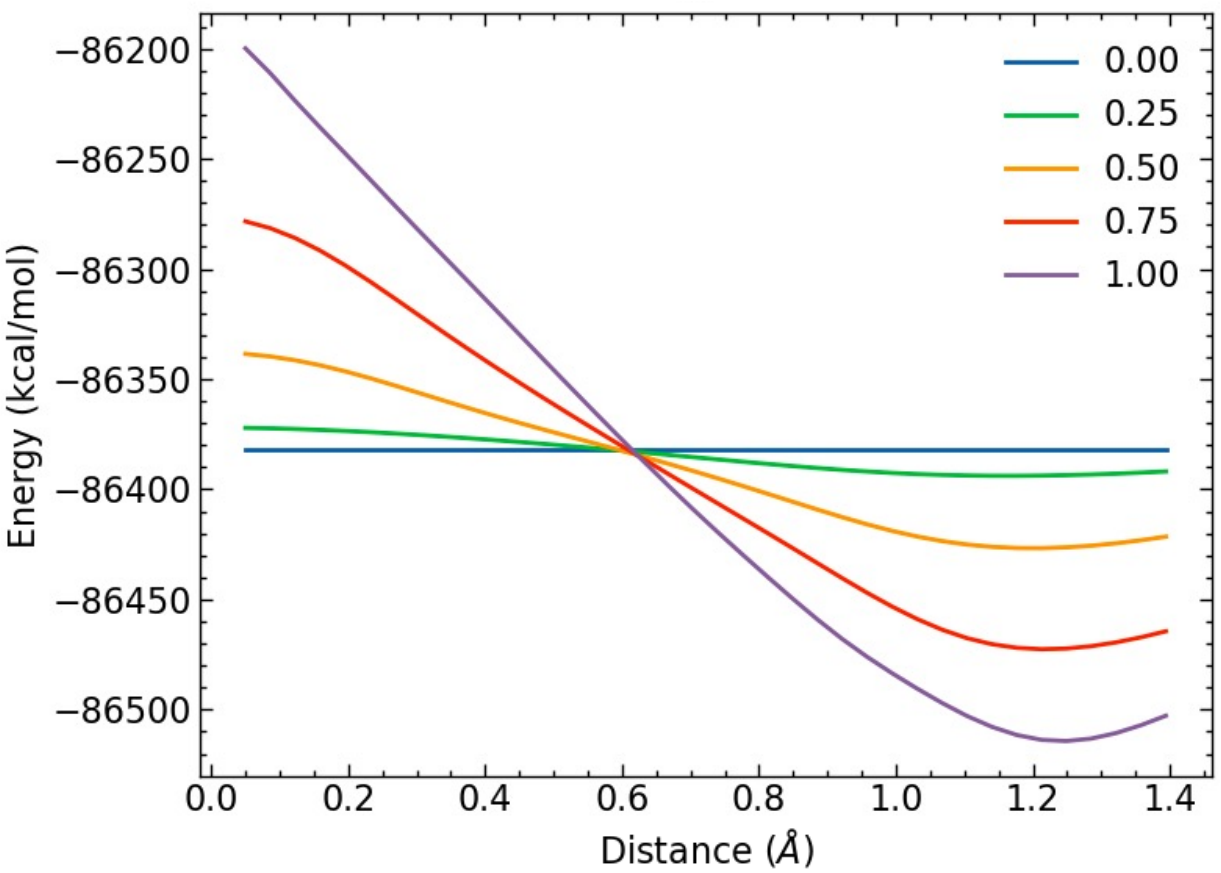}
    \caption{Dependence of softcore two-body interactions on $\lambda$ for a C--F diatomic pair learned by \texttt{MACE-OFF24-SC}.}
    \label{fig:lambda-dependent-softcore}
\end{figure}

In \texttt{MACE-OFF24-SC}, an approximately linear $\lambda$ schedule can be applied to the alchemical transformation (Figure~\ref{fig:lambda-dependent-softcore}), regardless of the size of the perturbation, compared to the highly skewed schedule required for linear softcore potentials~\cite{buelensLinearscalingSoftcoreScheme2012}. It is also worth noting that, unlike linear scaling schemes, this approach is amenable to nonequilibrium switching approaches, in which $\lambda$ is driven at a constant rate between the end states~\cite{gapsysLargeScaleRelative2020a, gapsysAccurateAbsoluteFree2021} (Section~\ref{SI:dhdl}).

While this approach can be applied in principle to any many-body potential that constructs site-wise energies as a tensor product of two-body features, we note that the success of this approach depends on the strong regularisation of MACE to enable smooth interpolation between the end states, as the contributions to the node energies are a nonlinear function of the (scaled) two-body features.

We also note that our alchemical modifications to the MACE graph share similarities with recently published work by Nam \textit{et al.}~\cite{namInterpolationDifferentiationAlchemical2024}. In particular, this work also introduces a scaling factor on the one-particle basis functions of MACE to compute free energy changes in solids. Whilst this is sufficient for studying crystalline systems, the additional inclusion of the softcore potential is crucial for stable alchemical condensed phase simulations, where, unlike materials applications, direct atomic overlap in partially decoupled states must be handled.

\subsection{Solvation free energy calculations}

Solvation free energies play a crucial role in assessing the accuracy of the intermolecular interactions within a forcefield and are an essential component of the standard evaluation of thermodynamic property prediction. These simulations probe the subtle nonbonded interactions between molecules, but their sampling requirements remain small relative to, for example, protein--ligand binding free energies, hence errors can be primarily attributed to the accuracy of the forcefield.

However, unlike properties such as densities and enthalpies of vapourisation and mixing, which are directly observable as ensemble averages from MD trajectories, free energies require evaluation of expectation values such as that shown in Eq.~\eqref{eq:thermoint}. Here we employ the MBAR estimator to compute the free energy difference between the end states A and B, by combining information from a series of intermediate alchemical windows in a statistically optimal fashion~\cite{bennettEfficientEstimationFree1976,shirtsStatisticallyOptimalAnalysis2008}. We stress, however, that our approach is agnostic to the estimator used.


These calculations can be efficiently performed using perturbations in parameter space, where the intermolecular interactions are gradually decoupled between the solute and solvent. Although such free energy calculations are routine with empirical forcefields, to the best of our knowledge, this work demonstrates the first rigorous, \textit{ab initio} quality solvation free energy calculations with all molecules treated using MLPs.  


\subsubsection{Hydration free energies}

\begin{figure}[H]
    \centering
    \includegraphics[width=1.0\linewidth]{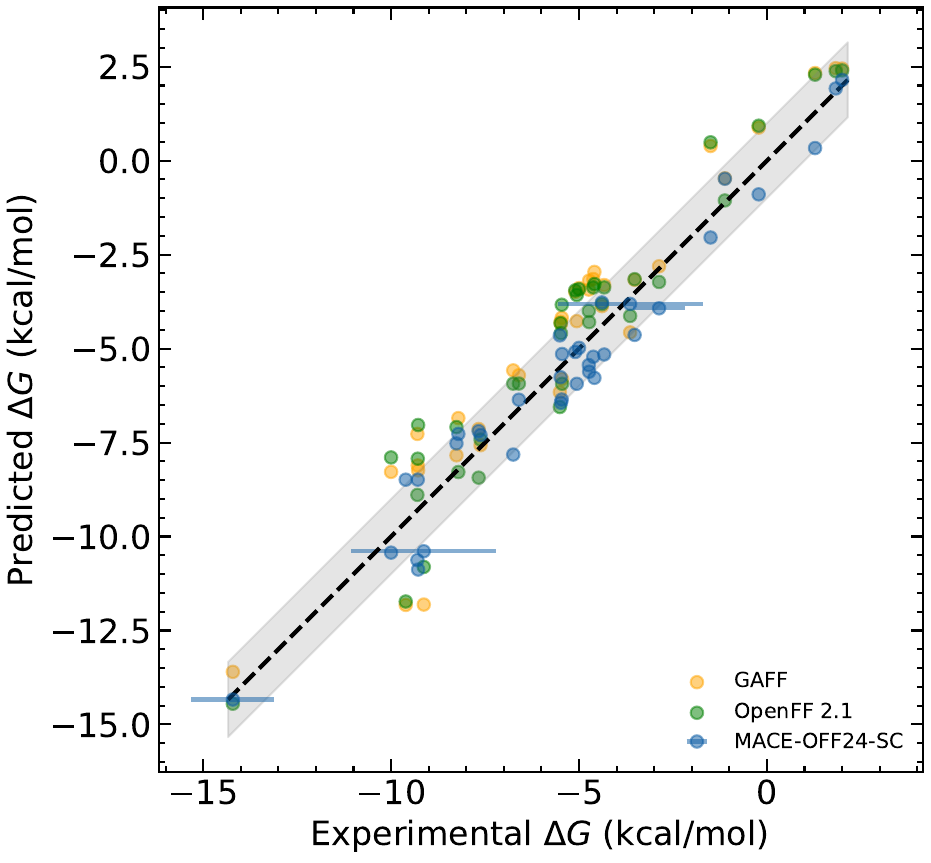}
    \caption{ Comparison of \texttt{MACE-OFF24-SC} hydration free energies with classical forcefields and experiment. Experimental error bars are shown for those compounds where the value exceeds 0.6 kcal/mol. The shaded region represents a 1~kcal/mol deviation.}
    \label{fig:hydr_correlation}
\end{figure}

We first evaluated the ability of \texttt{MACE-OFF24-SC} to compute accurate, converged hydration free energies for a series of organic molecules (Figure~\ref{fig:hydr_correlation}). 
For this test, we selected a diverse set of compounds from the FreeSolv database designed to cover a diverse range of functional groups relevant to medicinal chemistry~\cite{mobleyFreeSolvDatabaseExperimental2014}. Compounds were first binned according to their functional groups and subsequently filtered to remove those that were not unambiguously neutral at pH 7, as determined by Schr\"odinger's Epik program~\cite{epikx}. This ensured compatibility with \texttt{MACE-OFF24-SC}, which was trained on neutral molecules only. Compounds were then randomly sampled from each bin, ensuring good functional group coverage. The identities of the selected 36 compounds are tabulated in Section~\ref{si:freesolv}.

We performed replica exchange molecular dynamics with an ensemble of replicas with $\lambda$ windows spaced in the interval [0,1] to connect the physical end states.  To expedite phase space exploration,  exchanges of replicas between adjacent $\lambda$ windows were attempted every 1~ps.  

\begin{figure}[H]
    \centering
    \includegraphics[width=0.8\columnwidth]{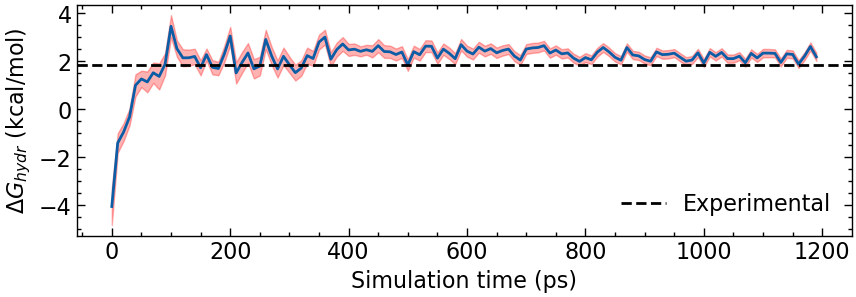}
    \caption{Convergence of ethane hydration free energy with simulation time.}
    \label{fig:ethane_solv_fe_timeseries}
\end{figure}

\begin{figure}[H]
    \centering
    \includegraphics[width=0.8\columnwidth]{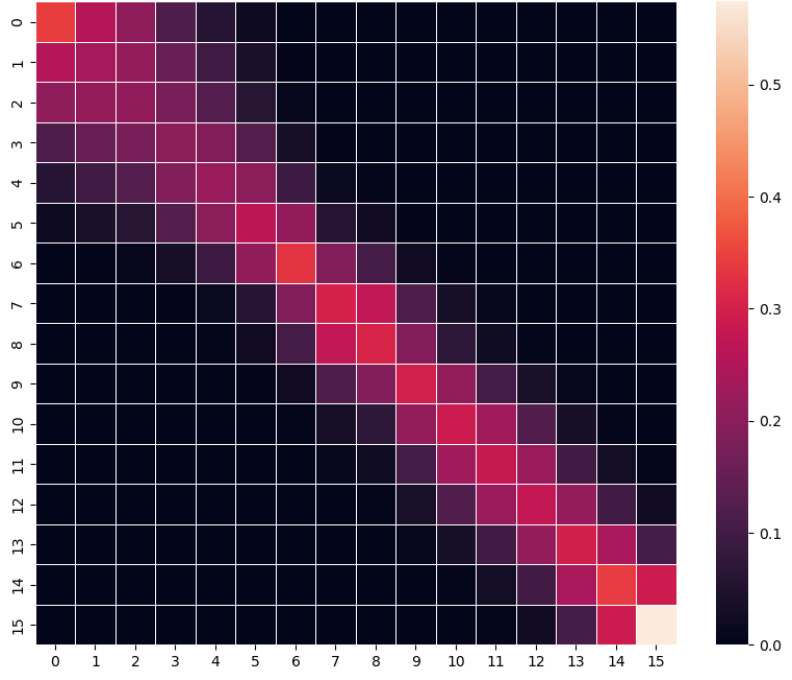}
    \caption{Transition probability matrix between 16 replicas from REMD simulation of phenol in water.}
    \label{fig:overlap_toluene}
\end{figure}

In all cases, simulations were performed with 16 replicas, which we found enabled sufficient phase-space overlap to converge the free energy simulations (Figure~\ref{fig:overlap_toluene}). The replica spacing was hand-tuned from an initial linear spacing to concentrate sampling in the region $0.15 \leq \lambda \leq 0.4$, where the curvature of $\frac{\partial H}{\partial \lambda}$ is greatest (Section~\ref{SI:dhdl}). For each system, convergence of the free energy as a function of time was confirmed (Figure~\ref{fig:ethane_solv_fe_timeseries} and Figure~\ref{SIfig:convergence_plots}).

Across the range of molecules studied here, we find that free energy predictions computed using \texttt{MACE-OFF24-SC} match or, in many cases, outperform widely-used classical forcefields, with a consistent improvement in MAE and RMSE compared to the classical forcefields (Table~\ref{tab:hydr_errors}).
\begin{table}[h!]
    \centering
    \begin{tabular}{@{}lcc@{}}
        \toprule
         & MAE & RMSE  \\
        \midrule
        MACE-OFF24-SC & 0.69 & 0.80  \\
        GAFF & 1.09 & 1.24  \\
        OpenFF 2.1 &  0.98 & 1.15   \\
        \bottomrule
    \end{tabular}
    \caption{Summary of hydration free energy errors (kcal/mol).}
        \label{tab:hydr_errors}
\end{table}

        
%
For example, compared to OpenFF~2.1~\cite{boothroydDevelopmentBenchmarkingOpen2023a} and GAFF2, we find that the \texttt{MACE-OFF24-SC} prediction for methanol falls within the experimental error, whilst the classical forcefields underpredict the magnitude of the hydration free energy.  This trend has previously been studied in the context of empirical forcefields and has been attributed to the assignment of partial charges on aliphatic hydroxyl groups~\cite{heFastHighqualityCharge2020, heABCG2MilestoneCharge2025}. Since our approach is not dependent upon the coarse graining of chemical space through atom-typing, our approach is not susceptible to the same inaccuracies.

Notwithstanding atom type specific parametrisation issues, it is encouraging to find that \texttt{MACE-OFF24-SC} is capable of exceeding the accuracy of classical forcefields, especially considering that the nonbonded parameters of both OpenFF 2.1 and GAFF2 have been parametrised to reproduce condensed phase properties of similar small molecules, whereas \texttt{MACE-OFF24-SC} must learn these interactions directly from the quantum mechanical training data. It is also worth noting that all forcefields are approaching the average experimental error of the dataset, around 0.6~kcal/mol, making further quantitative comparison challenging.




\subsubsection{Solvation free energies in a non-aqueous solvent}

We have shown that \texttt{MACE-OFF24-SC} is capable of highly accurate solvation free energy predictions across a range of small organic molecules. However, whilst these serve as a useful and important benchmark, often the more relevant quantity in drug discovery is the distribution coefficient, which additionally requires the calculation of a solvation free energy in a nonpolar solvent. This probes the ability of the model to describe accurately the intermolecular interactions between the solute and the nonpolar organic solvent, in this case octanol. Although the greater conformational flexibility of the solvent makes convergence of the octanol leg of the thermodynamic cycle slightly more challenging, these calculations were still performed in reasonable computational time, requiring at most 48 hours of walltime (see section~\ref{sec:comp_performance}). We first confirmed that \texttt{MACE-OFF24-SC} is capable of capturing the basic thermodynamic properties of liquid octanol, resulting a density within 5\% of the experimental value (Figure~\ref{SIfig:octanol_density}).

\begin{figure}
    \centering
    \includegraphics[width=1.0\linewidth]{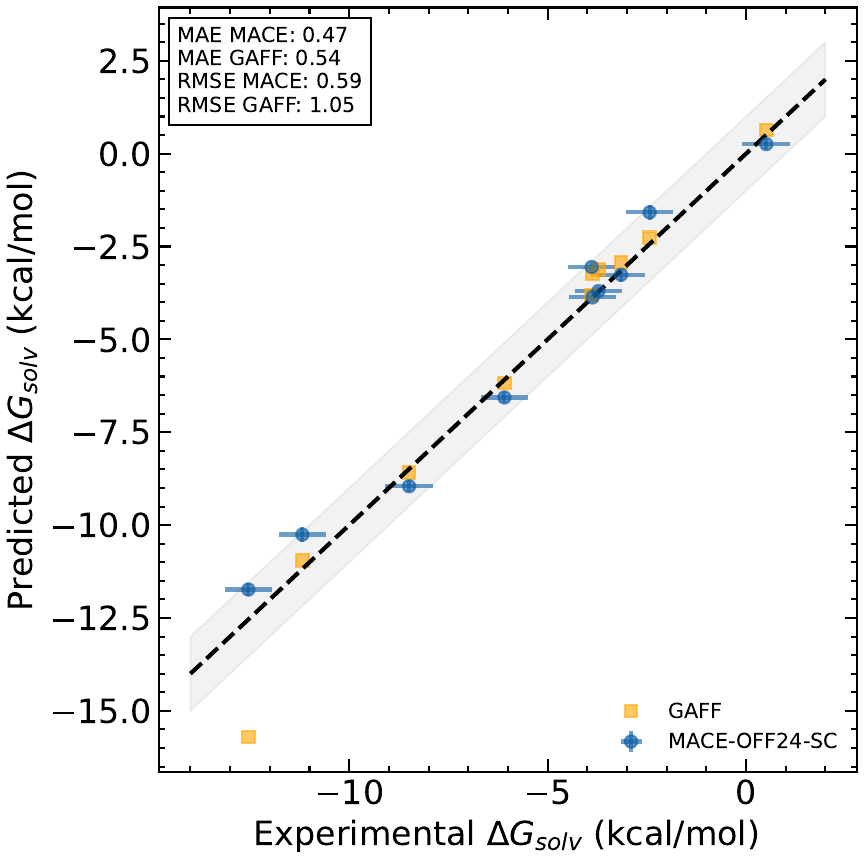}
    \caption{Solvation free energies of a series of 10 small organic molecules in octanol. The shaded region represents a 0.6~kcal/mol deviation.}
    \label{fig:octanol_solv}
\end{figure}

In this section, we evaluate the accuracy of \texttt{MACE-OFF24-SC} predictions on 10 octanol solvation free energies extracted from the MNSol database~\cite{marenichMinnesotaSolvationDatabase}. 
Whilst MNSol does not provide experimental uncertainties, we adopt the best practice developed by the FreeSolv authors and assume an experimental error of 0.6 kcal/mol. With this threshold, all \texttt{MACE-OFF24-SC} predictions fall within the experimental error (Figure~\ref{fig:octanol_solv} and Section~\ref{si:dgsolv}), indicating that the model is indeed capable of successfully predicting these interactions. Here, we compare to the GAFF2 forcefield equipped with the recently published ABCG2 charge model instead of the normal AM1-BCC, which has been tuned to improve the accuracy of hydration and transfer free energy calculations compared to the standard charge model~\cite{heABCG2MilestoneCharge2025}. Hence, this test represents an extremely stringent comparison for \texttt{MACE-OFF24-SC}, since the empirical forcefield has been directly optimised for the task.

Similar to the hydration free energy results, we find \texttt{MACE-OFF24-SC} has moderately lower prediction error compared to GAFF2/ABCG2, however both values fall below the presumed experimental error, again making further quantitative comparison difficult. 

\subsubsection{LogP calculations for drug-like molecules}

In order to probe the boundaries of the model's capabilities, we performed full solvation free energy calculations, in both water and octanol solvents, on a selection of 16 drug-like compounds from the CHEMBL database for which an experimental logP value is available via retention time-based UPLC/MS chromatographic analysis. All compounds satisfy Lipinski's Rule of 5 criteria~\cite{lipinskiLeadDruglikeCompounds2004} and are representative of the chemical complexity and flexibility of molecules typical of a medicinal chemistry campaign. Structures of the compounds are shown in Section~\ref{SI:chembl_compounds}.

Using the same alchemical free energy methodology as for the earlier calculations, we calculated logP as the difference between the water and octanol free energy legs:
%
\begin{align}
    \log P = \frac{\Delta G_{\text{hyd}} - \Delta G_{\text{oct}}}{2.303 RT}
    \label{eqn:logP}
\end{align}

\begin{figure*}
    \centering
    \includegraphics[width=0.8\linewidth]{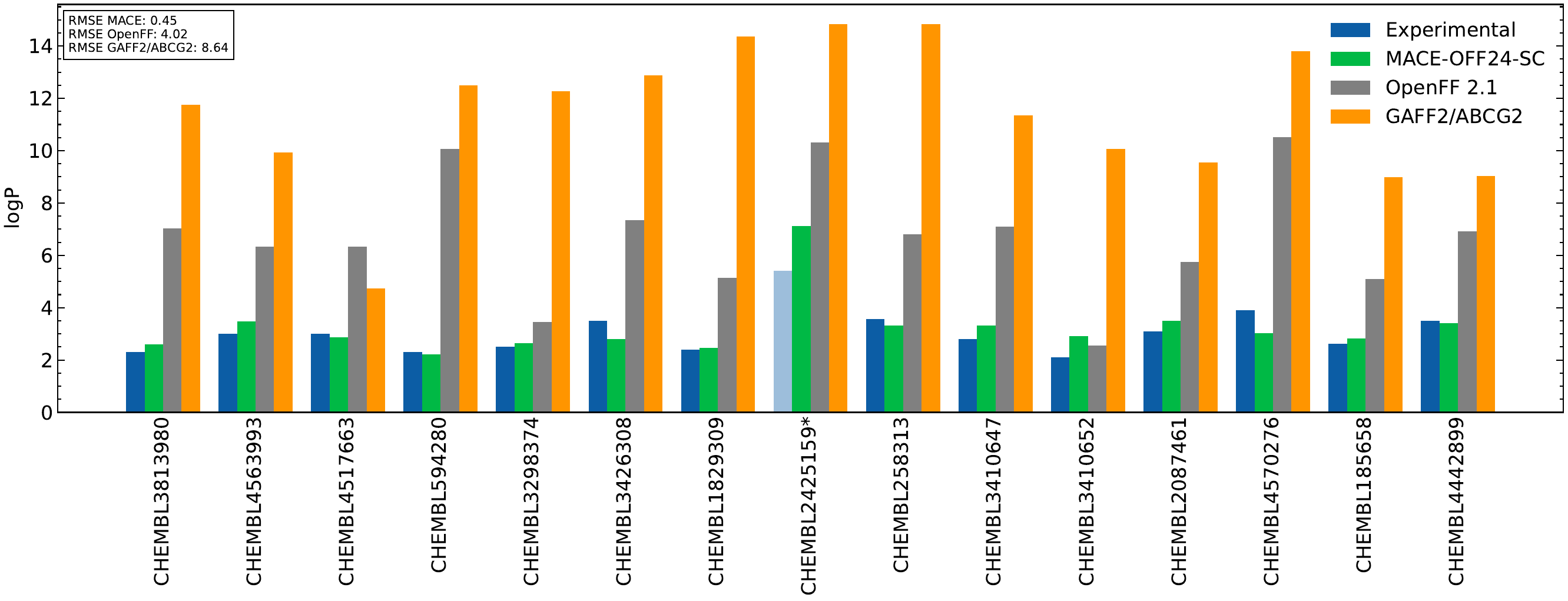}
    \caption{Experimental logP for CHEMBL compounds compared to OpenFF 2.1, \texttt{MACE-OFF24-SC} and GAFF2/ABCG2. \textsuperscript{*}Experimental value recorded as $>5.4$, so quantitative comparison is not possible.}
    \label{fig:chembl_logP}
\end{figure*}
Figure~\ref{fig:chembl_logP} compares both \texttt{MACE-OFF24-SC} and OpenFF~2.1 with experimental measurements.
Here, for the first time we see \texttt{MACE-OFF24-SC} significantly outperforming the classical forcefields, with an RMSE of 0.45 log units, compared to 4.02 for OpenFF 2.1 and 8.64 for GAFF2/ABCG2.  This result is extremely encouraging, suggesting that the model is capable of transferring nonbonded interactions learned from the training set to highly functionalised, conformationally complex molecules. This result validates our hypothesis that machine learning forcefields, which have learned these interactions from first principles, may show superior transferability to drug-like chemical space, compared to the explicit parametrisation approach used to fit empirical forcefields~\cite{haglerForceFieldDevelopment2019}.  Given the accurate predictions of GAFF2/ABCG2 on small molecule octanol solvation free energies, this suggests that the error arises in the transferability of the parameters from small molecules on which the nonbonded parameters are fitted to larger, more complex and conformationally flexible systems. These results are in qualitative agreement with other recently published benchmarking studies, which show that, whilst ABCG2 charges lead to improved performance on hydration free energies of small molecules, they fail to outperform AM1-BCC charges on closely related protein-ligand binding free energy calculations, further highlighting that parameter transferability remains a significant challenge for classical small molecule forcefields~\cite{beheraEvaluationABCG2Charge2025}. 

\subsubsection{Computational performance}
\label{sec:comp_performance}

Free energy calculations, even with classical forcefields, are known to be computationally expensive.  Hydration free energy calculations ameliorate the sampling problem for the purposes of benchmarking, since the phase space perturbations are small and expected to converge rapidly compared to protein-ligand binding free energies.  Running free energy calculations with machine learned potentials raises the issue of whether converged results can be obtained within reasonable wall time.

Here, all calculations were performed on single nodes, containing either 8 NVIDIA A100, or 8 NVIDIA L40S, GPUs. For both machine types, we achieved aggregated sampling of around 8~ns/day across 16 replicas, corresponding to approximately 500 replica exchange iterations of 1~ps each per day. Most hydration free energies converged within this time, whilst some were extended to run for 1~ns per replica, requiring 48 hours of wall time. MNSol octanol solvation free energies were all run for 1~ns per replica to account for the additional conformational sampling requirement.

Owing to the size of the solutes in the logP calculations, these calculations took longer to converge, with the average hydration and octanol leg requiring 4 and 7 days of walltime, respectively.

\section{Conclusion}

We have introduced a modified version of the \texttt{MACE-OFF24} transferable organic MLPs that enables, theoretically rigorous condensed phase alchemical free energy simulations with first principles based MLPs.  A dual approach of softcore repulsion and $\lambda$-dependent two-body interactions enable numerically well-behaved simulations of weakly interacting atoms, directly analogous to softcore formulations of classical forcefields.  Our method enables access to quantities directly relevant to drug discovery at \textit{ab initio} quality.

We tested the accuracy of our approach by calculating free energies of solvation, in water and octanol solvents, for a range of small organic molecules from the FreeSolv and MNSol databases~\cite{mobleyFreeSolvDatabaseExperimental2014, marenichMinnesotaSolvationDatabase}.  Our approach achieves excellent accuracy on a wide range of functional groups, improving on state-of-the-art empirical forcefield accuracy and approaching the experimental errors of the datasets. 

As a final stringent test of the accuracy of the model, we computed logP values for a series of compounds from the CHEMBL database. These compounds are drug-like in terms of size, complexity and functionality, are expected to be neutral at physiological pH and have experimental logP values. 
On this benchmark, we report an order of magnitude improvement in RMSE compared to OpenFF 2.1, a state-of-the-art classical forcefield, and GAFF2/ABCG2, which utilises a recently developed charge partitioning scheme~\cite{heABCG2MilestoneCharge2025}. This performance is highly encouraging for future applications of machine learning forcefields in drug discovery, and represents a significant step towards replacing labour and resource intensive chemical synthesis and experiment with fast and accurate molecular simulation.

Equally importantly, we demonstrate that convergence of the free energy is feasible within reasonable simulation time, even for absolute solvation free energy calculations in octanol, which present additional sampling challenges compared to aqueous free energy calculations. Since this work was performed, optimised computational kernels such as NVIDIA's cuEquivariance have been released\footnote{https://github.com/NVIDIA/cuEquivariance}, which we expect to significantly improve the throughput of these calculations.

As has been previously noted for MACE-OFF models, the lack of an explicit long-range contribution limits the chemical space in which the model is expected to be accurate. This has been addressed by recent transferable MLPs including AIMNet2~\cite{anstineAIMNet2NeuralNetwork2024} and FeNNix-Bio1~\cite{pleFoundationModelAccurate2025a} which include an explicit Coulomb term. The latter employs the alchemical methods developed in this work and uses solvation and protein-ligand binding free energies to benchmark the accuracy of the model for condensed phase systems.

We believe softcore-equipped machine learning potentials will become an increasingly exploited avenue for accurate and efficient calculation of free energies.  As well as being an important quantity in forcefield benchmarking, free energies represent key optimisation quantities in drug discovery, where they can give direct access to protein--ligand binding affinities. Hence accurate and efficient computational methods are an important tool when rigorous and accurate predictions are required. 


\section{Methods}

\subsection{\texttt{MACE-OFF24-SC} training}

In previous work, it was found that the use of equivariant features ($L_{max} > 0$) in the tensor product significantly increased the accuracy of intermolecular force predictions compared to the invariant MACE-OFF23 model~\cite{kovacsMACEOFFShortRangeTransferable2025}.  All simulations in this work were performed using a model with $L_{max}=1$,  128 channels, and a layer-wise cutoff of 6~\AA{}. These hyperparameters were adopted from the \texttt{MACE-OFF24(M)} model, given its ability to accurately describe the properties of condensed phase systems~\cite{kovacsMACEOFFShortRangeTransferable2025}.

\texttt{MACE-OFF24-SC} was trained on the same neutral-only filtering of the SPICE2 dataset that we previously reported, which included, crucially, the newly added protein-ligand and solvated pubchem datasets, which provided extensive sampling of the intermolecular interactions required for accurate liquid phase free energy calculations~\cite{eastmanSPICE2012024}.

\subsection{Equilibrium free energy calculations}
Equilibrium free energy calculations were performed using OpenMM, and made use of modified versions of the \texttt{openmmtools}, \texttt{openmm-ml} and \texttt{openmm-torch} libraries to perform Hamiltonian replica exchange and interface MACE with OpenMM~\cite{eastmanOpenMMMolecularDynamics2023,choderaReplicaExchangeExpanded2011a,eastmanConstantConstraintMatrix2010, eastmanEfficientNonbondedInteractions2010, eastmanOpenMMHardwareIndependentFramework2010, friedrichsAcceleratingMolecularDynamic2009}. This approach enabled interpolation between the coupled and decoupled endstates via a $\lambda$-dependent Hamiltonian.\footnote{https://github.com/jharrymoore/mace-md}

Since MACE predicts total intermolecular interactions without distinguishing between repulsive exchange and attractive Coulomb and dispersion terms, solvent-solute interactions are switched off using a single transformation, parameterised by the alchemical parameter $\lambda$. We adopt a direct decoupling approach in which the organic molecule is decoupled from the surrounding solvent, whilst all other interactions are unmodified. This results in a decoupled end state equivalent to a separated solvent box and small molecule in vacuum.

Initial solvated structures were generated with \texttt{pdbfixer}, using a rhombic dodecahedral box and a padding of 1.3~nm from solute to box edge, ensuring the minimum image convention is correctly applied for the 12~\AA~receptive field of MACE.  Structures were energy minimised using the L-BFGS algorithm, and subsequently equilibrated in the NPT ensemble for 50 ps. The last frame was used to seed the replica exchange calculations.

Replicas were then propagated in the NPT ensemble under Langevin dynamics with an integration timestep of 1~fs and a friction coefficient of 1/ps. Pressure was maintained at 1 atm with a MonteCarlo barostat as implemented in OpenMM.

The phase space perturbation was spanned in all cases by 16 replicas, which resulted in at least tridiagonal overlap in all cases. Hamiltonian replica exchange between all replicas were attempted every 1000 steps of dynamics (1~ps). It was found that 1~ns of sampling per replica was sufficient to converge the free energy estimates in all cases. A full sample input file is provided in Section~\ref{SIsec:repex_input}

Replica exchange calculations were performed using \texttt{ReplicaExchangeSampler} class from \texttt{openmmtools}.   
Replicas were assigned a single MPI rank, and two MPI ranks were mapped to each GPU, enabling full utilisation of the GPU's compute bandwidth, thus slightly increasing throughput compared to running the replicas in multiple waves.

Automated equilibration detection and trajectory subsampling were performed following Chodera \textit{et al.} as implemented in \texttt{pymbar} to ensure only decorrelated samples were used to solve the MBAR equations~\cite{choderaSimpleMethodAutomated2016, shirtsStatisticallyOptimalAnalysis2008}.

\subsection{Empirical forcefield solvation free energies}

GAFF solvation free energies and standard errors were taken from the FreeSolv Dataset~\cite{mobleyFreeSolvDatabaseExperimental2014}.  OpenFF values were calculated using OpenFF 2.1.0 via the \texttt{Absolv} code, using an equilibrium free energy protocol\footnote{https://github.com/SimonBoothroyd/absolv/}. An example script can be found in Section~\ref{SIsec:absolv}.


GAFF2/ABCG2 calculations were performed with GROMACS 2024.1. Parameters were assigned with Antechamber. Systems were energy minimised for a maximum of 50000 steps and subsequently equilibrated for 1 ns in the NVT ensemble and the NPT ensemble for 1 ns. 8 and 16 windows were used to annihilate the Coulomb and Van der Waals interactions respectively. Each alchemical window was run for 500~ps. An example MDP file for the production free energy calculations is included in the section \ref{SIsec:gmx_mdp}. Free energy estimates were calculated using \texttt{gmx bar}.


\section*{Acknowledgements}
JHM acknowledges support from an AstraZeneca Non-Clinical PhD studentship, and thanks Prof. Ola Engkvist, Dr Marco Kl\"ahn and Dr Graeme Robb for their helpful discussions. DJC acknowledges support from a UKRI Future Leaders Fellowship (grant MR/T019654/1). We acknowledge computational resources provided by the Cambridge Service for Data-Driven Discovery (CSD3), the AstraZeneca Scientific Computing Platform (SCP), the UK national high-performance computing service, ARCHER2, for which access was obtained via the UKCP consortium and the EPSRC grant ref EP/P022561/1, and Ångström AI.

\section*{Competing interests}
GC has equity stake in Symmetric Group LLP. GC and JHM have equity stake in \AA{}ngstr\"om AI Inc. Both companies are engaged in the application of machine learning to material and molecular simulation.

\section*{Data Availability}

An example script to reproduce the replica exchange calculations is provided in the SI. The \texttt{mace-md} package used to run the simulations is available at \texttt{https://github.com/jharrymoore/mace-md/tree/master}. Calculated solvation free energy values for all forcefields are available in the supplementary information.

\section*{Supplementary Information}

The Supplemetary information contains the following sections
\begin{enumerate}
    \item Training and test set energy and force errors
    \item Octanol solvation free energy results
    \item FreeSolv hydration free energy results
    \item Switching points
    \item Example of nonequilibrium switching
    \item Free energy convergence plots
    \item Sample replica exchange input file
    \item Example Absolv run script
    \item CHEMBL compounds
    \item Octanol Density
    \item Example GROMACS MDP file
\end{enumerate}

\bibliography{refs}





\appendix
\begin{widetext}

\input{SI}

\end{widetext}

\end{document}

%% file: SI.tex
\newcommand*{\mycommand}[1]{\texttt{\emph{#1}}}
\newcommand*\lm[1]{{\color{red}\tt#1}}
\newcommand*\csg[1]{{\color{blue}\tt#1}}

\renewcommand{\thefigure}{S\arabic{figure}}
\renewcommand{\thesection}{S\arabic{section}}
\renewcommand{\thetable}{S\arabic{table}}
\renewcommand\refname{Supplementary References}

\definecolor{codegreen}{rgb}{0,0.6,0}
\definecolor{codegray}{rgb}{0.5,0.5,0.5}
\definecolor{codepurple}{rgb}{0.58,0,0.82}
\definecolor{backcolour}{rgb}{0.95,0.95,0.92}

\lstdefinestyle{bashstyle}{
    backgroundcolor=\color{backcolour},   
    commentstyle=\color{codegreen},
    keywordstyle=\color{magenta},
    numberstyle=\tiny\color{codegray},
    stringstyle=\color{codepurple},
    basicstyle=\ttfamily\footnotesize,
    breakatwhitespace=false,         
    breaklines=true,                 
    captionpos=b,                    
    keepspaces=true,                 
    numbers=left,                    
    numbersep=5pt,                  
    showspaces=false,                
    showstringspaces=false,
    showtabs=false,                  
    tabsize=2,
    frame=single,
    rulecolor=\color{black}
}

\lstset{style=bashstyle, language=bash}

\title{Supporting Information for: \\Computing solvation free energies of small molecules with experimental accuracy}

\author{J. Harry Moore\textsuperscript{\dag}}
\affiliation{Engineering Laboratory, University of Cambridge, Cambridge, CB2 1PZ, UK}
\affiliation{Ångström AI, 2325 3rd Street, San Francisco, CA 94107, USA}
\author{Daniel J. Cole}
\affiliation{School of Natural and Environmental Sciences, Newcastle University, Newcastle upon Tyne NE1 7RU, UK}

\author{G\'abor Cs\'anyi}
\affiliation{Engineering Laboratory, University of Cambridge, Cambridge, CB2 1PZ, UK}
\affiliation{Ångström AI, 2325 3rd Street, San Francisco, CA 94107, USA}

\email{jhm72@cam.ac.uk}


\clearpage


\section{Training and test set energy and force errors}

\label{sec:test_errors}
\begin{table}[H]
  \centering
  \caption{Route mean squared errors (RMSE) for training, validation and test set configurations for \texttt{MACE-OFF24-SC}. See previous work for details of the datasets~\cite{kovacsMACEOFFShortRangeTransferable2025}. Energy errors are in meV/atom, and force errors in meV/\AA{}.}
    \begin{tabular}{lccc}
    \toprule
    \textbf{Dataset type} & \textbf{E RMSE~~} & \textbf{F RMSE~~} & \textbf{Rel. F RMSE \%} \\
    \midrule
    Training & 1.2   & 28.7  & 3.96 \\
    Validation & 1.2   & 28.9  & 3.70 \\
    DES370K Dimers & 2.1   & 15.2  & 3.23 \\
    DES370K Monomers & 2.5   & 16.4  & 2.54 \\
    Dipeptides & 2.5   & 22.9  & 3.25 \\
    PubChem & 2.8   & 33.7  & 4.36 \\
    QMugs & 2.3   & 35.5  & 3.53 \\
    Solvated Amino Acids & 1.3   & 33.4  & 2.73 \\
    Water & 1.9   & 23.6  & 3.72 \\
    \bottomrule
    \end{tabular}%
  \label{tab:addlabel}%
\end{table}%

\clearpage
\section{Octanol solvation free energy results}
\label{si:dgsolv}

\begin{table}[H]
\centering
\tiny
\begin{tabular}{llccccc}
\toprule
\textbf{Molecule} & \textbf{SMILES} & \textbf{Experiment} & \textbf{Experiment Error} & \textbf{MACE} & \textbf{MACE Error} & \textbf{GAFF2/ABCG2} \\
\midrule
methanol     & \texttt{CO}                         & -3.87  & 0.6 & -3.87  & 0.09 & -3.223  \\
phthalimide  & \texttt{O=C1NC(=O)c2ccccc12}        & -11.18 & 0.6 & -10.25 & 0.21 & -10.945 \\
methane      & \texttt{C}                          & 0.51   & 0.6 & 0.26   & 0.15 & 0.64    \\
thiophene    & \texttt{c1ccsc1}                    & -3.89  & 0.6 & -3.05  & 0.11 & -3.807  \\
benzonitrile & \texttt{N\#Cc1ccccc1}               & -6.09  & 0.6 & -6.56  & 0.15 & -6.169  \\
caffeine     & \texttt{Cn1cnc2c1c(=O)n(C)c(=O)n2C} & -12.54 & 0.6 & -11.73 & 0.17 & -15.708 \\
bromomethane & \texttt{CBr}                        & -2.43  & 0.6 & -1.58  & 0.2  & -2.264  \\
o-cresol     & \texttt{Cc1ccccc1O}                 & -8.49  & 0.6 & -8.95  & 0.16 & -8.58   \\
benzene      & \texttt{c1ccccc1}                   & -3.72  & 0.6 & -3.7   & 0.11 & -3.121  \\
acetone      & \texttt{CC(=O)C}                    & -3.15  & 0.6 & -3.26  & 0.19 & -2.935  \\
\bottomrule
\end{tabular}
\caption{Solvation free energies in kcal/mol in octanol for a series of organic solutes. GAFF2-ABCG2 values are taken from the literature.~\cite{heABCG2MilestoneCharge2025}}
\end{table}

\clearpage

\section{FreeSolv hydration free energy results}
\label{si:freesolv}

\begin{table}[H]
\centering
\tiny
\begin{tabular}{lcccccccc}
\toprule
Compound Name & Exp & MACE & GAFF & OpenFF & Exp Error & MACE Error & GAFF Error & OpenFF Error \\
\midrule
cycloheptanol & -5.48 & -6.44 & -4.34 & -4.59 & 0.60 & 0.18 & 0.03 & 0.19 \\
ethylene & 1.28 & 0.33 & 2.33 & 2.29 & 0.60 & 0.09 & 0.01 & 0.09 \\
2-chloropyridine & -4.39 & -3.77 & -3.87 & -3.81 & 0.60 & 0.14 & 0.03 & 0.16 \\
2-methylpyrazine & -5.51 & -4.65 & -6.16 & -6.55 & 0.60 & 0.15 & 0.03 & 0.18 \\
pentan-2-one & -3.52 & -4.64 & -3.17 & -3.15 & 0.60 & 0.15 & 0.03 & 0.20 \\
3-ethylpyridine & -4.59 & -5.78 & -2.96 & -3.28 & 0.60 & 0.24 & 0.03 & 0.17 \\
diphenyl ether & -2.87 & -3.93 & -2.81 & -3.23 & 0.69 & 0.27 & 0.03 & 0.19 \\
naphthalen-1-ol & -7.67 & -7.20 & -7.14 & -8.43 & 0.60 & 0.19 & 0.03 & 0.20 \\
fluoromethane & -0.22 & -0.90 & 0.88 & 0.93 & 0.60 & 0.13 & 0.01 & 0.10 \\
phthalimide & -9.61 & -8.49 & -11.82 & -11.73 & 0.50 & 0.15 & 0.03 & 0.22 \\
propan-2-ol & -4.74 & -5.44 & -3.43 & -4.00 & 0.60 & 0.14 & 0.02 & 0.16 \\
pyridine-3-carbonitrile & -6.75 & -7.82 & -5.58 & -5.93 & 0.60 & 0.17 & 0.03 & 0.17 \\
1-(3-pyridyl)ethanone & -8.26 & -7.52 & -7.84 & -7.09 & 0.60 & 0.20 & 0.03 & 0.20 \\
cyclopentanol & -5.49 & -5.76 & -4.29 & -4.32 & 0.60 & 0.15 & 0.03 & 0.17 \\
4-ethylpyridine & -4.73 & -5.62 & -3.19 & -4.29 & 0.60 & 0.15 & 0.03 & 0.17 \\
1-(4-pyridyl)ethanone & -7.62 & -7.30 & -7.57 & -7.42 & 0.60 & 0.32 & 0.03 & 0.21 \\
methane & 2.00 & 1.33 & 2.45 & 2.40 & 0.20 & 0.09 & 0.01 & 0.15 \\
ethanol & -5.00 & -5.83 & -3.39 & -4.98 & 0.60 & 0.16 & 0.02 & 0.12 \\
phenol & -6.60 & -6.29 & -5.71 & -6.36 & 0.20 & 0.13 & 0.03 & 0.15 \\
ethane & 1.83 & 1.02 & 2.46 & 1.92 & 0.60 & 0.12 & 0.01 & 0.14 \\
methanol & -5.10 & -5.61 & -3.49 & -5.09 & 0.60 & 0.16 & 0.02 & 0.17 \\
N,N-dimethylbenzamide & -9.29 & -8.49 & -8.11 & -7.93 & 0.60 & 0.27 & 0.03 & 0.21 \\
N-methylacetamide & -10.00 & -10.43 & -8.28 & -7.90 & 0.60 & 0.31 & 0.02 & 0.18 \\
ethylene glycol & -9.30 & -10.63 & -7.27 & -8.89 & 0.60 & 0.15 & 0.03 & 0.20 \\
alachlor & -8.21 & -7.27 & -6.85 & -8.28 & 0.29 & 0.26 & 0.05 & 0.25 \\
chlorobenzene & -1.12 & -0.48 & -0.47 & -1.06 & 0.60 & 0.15 & 0.02 & 0.15 \\
methylsulfinylmethane & -9.28 & -10.88 & -8.24 & -7.03 & 0.57 & 0.20 & 0.02 & 0.17 \\
2-ethylpyridine & -4.33 & -5.16 & -3.31 & -3.37 & 0.60 & 0.16 & 0.03 & 0.19 \\
fenuron & -9.13 & -10.39 & -11.81 & -10.81 & 1.93 & 0.17 & 0.04 & 0.23 \\
2-ethylpyrazine & -5.45 & -5.14 & -5.81 & -5.94 & 0.60 & 0.19 & 0.03 & 0.21 \\
pebulate & -3.64 & -3.81 & -4.57 & -4.13 & 1.93 & 0.24 & 0.04 & 0.23 \\
\makecell[l]{1-(2-hydroxyethylamino)-\\9,10-anthraquinone} & -14.21 & -14.34 & -13.60 & -14.45 & 1.10 & 0.34 & 0.05 & 0.28 \\
cyclohexanol & -5.46 & -6.35 & -4.18 & -3.83 & 0.60 & 0.14 & 0.03 & 0.21 \\
1,4-dioxane & -5.06 & -5.94 & -4.27 & -3.58 & 0.60 & 0.16 & 0.02 & 0.16 \\
butan-2-ol & -4.62 & -5.22 & -3.15 & -3.37 & 0.60 & 0.18 & 0.03 & 0.17 \\
methylsulfanylethane & -1.5 & -2.04 & 0.39 & 0.49 & 0.2 & 0.15 & 0.02 & 0.14 \\
\bottomrule
\end{tabular}
\caption{Solvation free energies in kcal/mol for MACE, OpenFF and GAFF on the FreeSolv subset compared to experiment.}
\end{table}


\section{Switching points}
\label{SI:switching_points}

\begin{table}[H]
\centering
\begin{tabular}{cc@{\hskip 2cm}cc}

\toprule
\textbf{Dimer pair} & \textbf{Switching point / \AA{}} & \textbf{Dimer pair} & \textbf{Switching point / \AA{}} \\
\midrule
Br2        & 2.1499996        & FO         & 1.2799997        \\
BrCl       & 2.0199995        & FP         & 1.5199996        \\
BrF        & 1.6749996        & FS         & 1.5249996        \\
BrI        & 2.3149996        & H2         & 0.6749997        \\
BrN        & 1.6399996        & HBr        & 1.3299996        \\
BrO        & 1.6299996        & HCl        & 1.2049997        \\
BrP        & 2.0249996        & HF         & 0.87999976       \\
BrS        & 2.0049996        & HI         & 1.4949995        \\
C2         & 1.2149997        & HN         & 0.9749997        \\
CBr        & 1.7049996        & HO         & 0.9199998        \\
CCl        & 1.5549996        & HP         & 1.3099995        \\
CF         & 1.2149997        & HS         & 1.2499996        \\
CH         & 1.0349996        & I2         & 2.480009         \\
CI         & 1.8849996        & IN         & 1.8049996        \\
CN         & 1.1349998        & IO         & 1.7749996        \\
CO         & 1.0999999        & IP         & 2.1949997        \\
CP         & 1.4949996        & IS         & 2.1649995        \\
CS         & 1.4749997        & N2         & 1.0699998        \\
Cl2        & 1.8849995        & NO         & 1.1149998        \\
ClF        & 1.5499996        & NP         & 1.4349997        \\
ClI        & 2.1799996        & NS         & 1.4399997        \\
ClN        & 1.4899997        & O2         & 1.1649997        \\
ClO        & 1.4899997        & OP         & 1.4299997        \\
ClP        & 1.8899995        & OS         & 1.4249997        \\
ClS        & 1.8749995        & P2         & 1.8049997        \\
F2         & 1.3249997        & PS         & 1.8149996        \\
FI         & 1.8149996        & S2         & 1.8049996        \\
FN         & 1.2399998        &            &                  \\
\bottomrule
\end{tabular}
\end{table}

\section{Example of nonequilibrium switching}
\label{SI:dhdl}

\begin{figure}[H]
    \centering
    \includegraphics[width=0.5\linewidth]{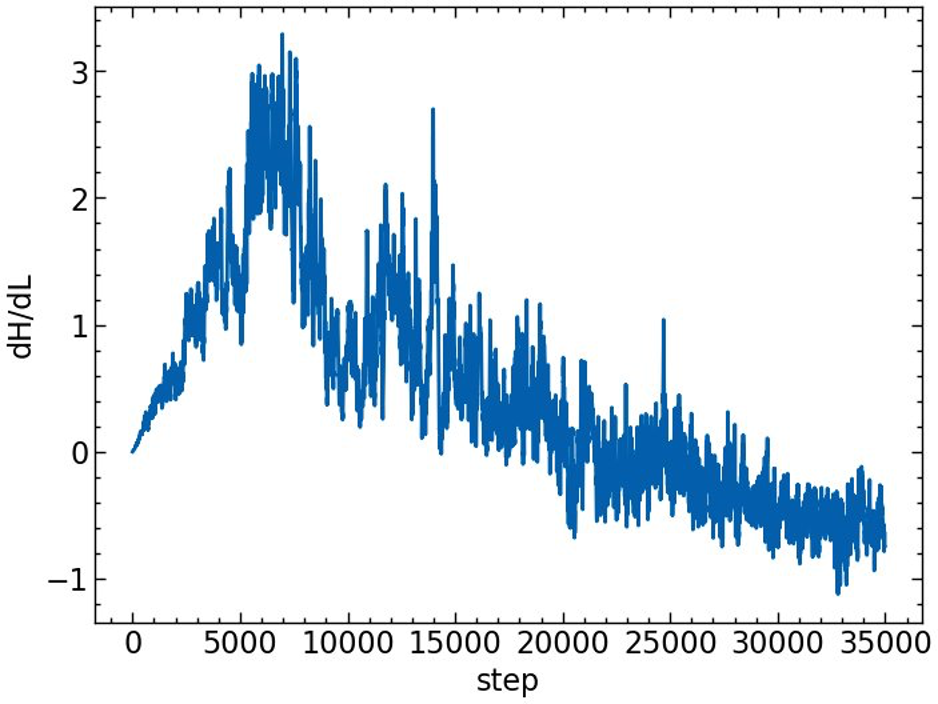}
    \caption{Example $\frac{dH}{d\lambda}$ plot for the solvation of methane in water over a 35~ps nonequilibrium switching trajectory.}
\end{figure}

\section{Free energy convergence plots}
\begin{figure}[H]
    \centering
    \includegraphics[width=0.5\linewidth]{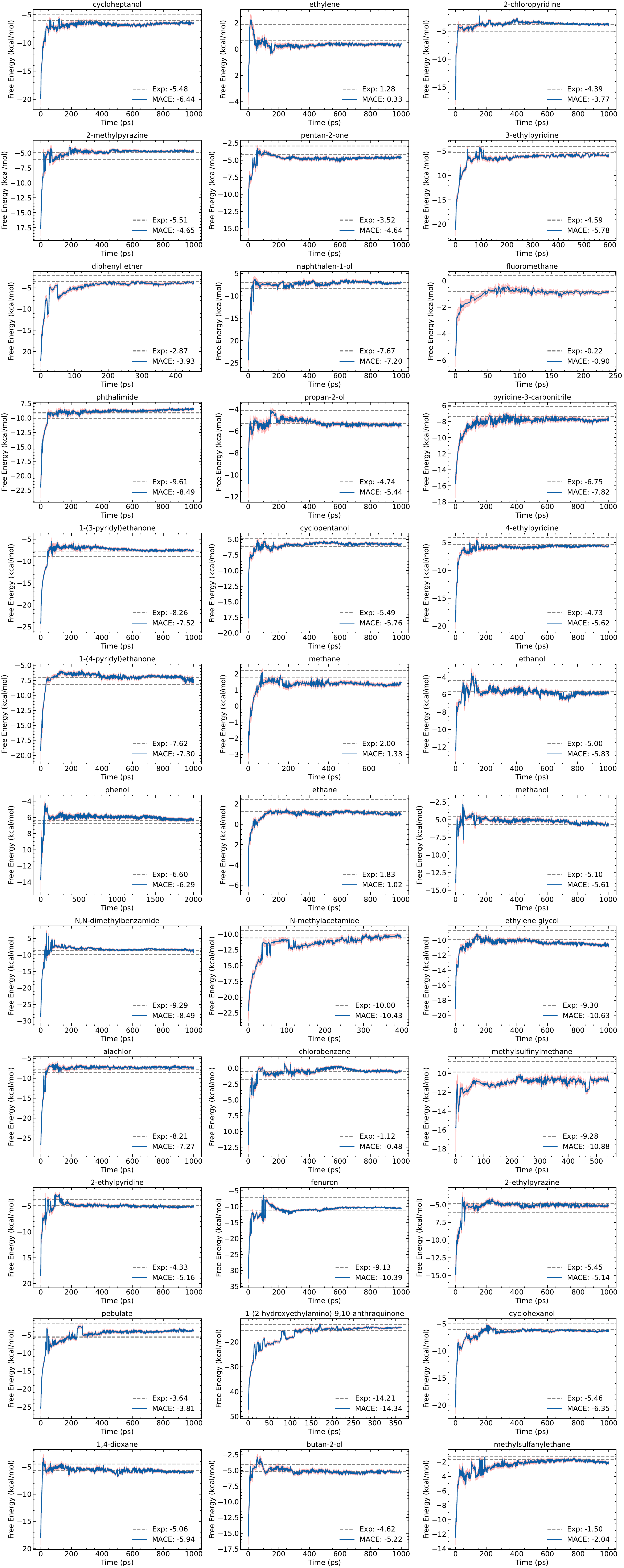}
    \caption{Convergence of hydration free energy calculations from FreeSolv benchmark.}
    \label{SIfig:convergence_plots}
\end{figure}

\section{Sample replica exchange input file}
\label{SIsec:repex_input}

\begin{lstlisting}[caption=Script to run 1 ns of replica exchange simulation via the MACE-MD code]
#!/bin/bash
mpirun -np 8 mace-md -f /path/to/equilibrated/structure.pdb \
  --resname "LIG"\
  --output_dir /path/to/output/dir \
  --pressure 1.0 \
  --decouple \
  --restart \
  --dtype float32 \
  --log_level INFO \
  --run_type "repex" \
  --system_type "pure" \
  --interval 1 \
  --steps 1000 \
  --lambda_schedule="[0.  , 0.08, 0.1, 0.13,  0.15, 0.17, 0.20, 0.25, 0.2825, 0.385 , 0.4875, 0.59  , 0.6925, 0.795 , 0.8975, 1.]"\
  --model_path "/path/to/mace/model" \
  --steps_per_iter 1000 \
\end{lstlisting}
\clearpage

\section{Example Absolv run script}
\label{SIsec:absolv}

\begin{lstlisting}[caption=Script to run equilibrium hydration free energy calculation with openff-2.1.0.]
import pathlib
import sys
import openmm.unit
import openff.toolkit
from absolv import config, runner
import absolv

smiles = sys.argv[1]
temperature=298.15 * openmm.unit.kelvin
pressure=1.0 * openmm.unit.atmosphere

system=absolv.config.System(
    solutes={smiles: 1}, solvent_a=None, solvent_b={"O": 1500}
)

alchemical_protocol_a=absolv.config.EquilibriumProtocol(
    lambda_sterics=[
        1.00, 1.00, 1.00, 1.00, 1.00, 0.95, 0.90, 0.80, 0.70, 0.60, 0.50, 0.40,
        0.35, 0.30, 0.25, 0.20, 0.15, 0.10, 0.05, 0.00,
    ],
    lambda_electrostatics=[
        1.00, 0.75, 0.50, 0.25, 0.00, 0.00, 0.00, 0.00, 0.00, 0.00, 0.00, 0.00,
        0.00, 0.00, 0.00, 0.00, 0.00, 0.00, 0.00, 0.00,
    ]
)
alchemical_protocol_b=absolv.config.EquilibriumProtocol(
    lambda_sterics=[
        1.00, 1.00, 1.00, 1.00, 1.00, 0.95, 0.90, 0.80, 0.70, 0.60, 0.50, 0.40,
        0.35, 0.30, 0.25, 0.20, 0.15, 0.10, 0.05, 0.00,
    ],
    lambda_electrostatics=[
        1.00, 0.75, 0.50, 0.25, 0.00, 0.00, 0.00, 0.00, 0.00, 0.00, 0.00, 0.00,
        0.00, 0.00, 0.00, 0.00, 0.00, 0.00, 0.00, 0.00,
    ]
)

config = absolv.config.Config(
    temperature=temperature,
    pressure=pressure,
    alchemical_protocol_a=alchemical_protocol_a,
    alchemical_protocol_b=alchemical_protocol_b,
)

force_field = openff.toolkit.ForceField("openff-2.1.0.offxml")
prepared_system_a, prepared_system_b = absolv.runner.setup(system, config, force_field)

result = absolv.runner.run_eq(
    config, prepared_system_a, prepared_system_b, "CUDA"
)

path = pathlib.Path("results")
(path / "output.txt").write_text(result.model_dump_json(indent=2))
\end{lstlisting}
\clearpage

\section{CHEMBL compounds}
\label{SI:chembl_compounds}

\begin{figure}[H]
    \centering
    \includegraphics[width=0.8\linewidth]{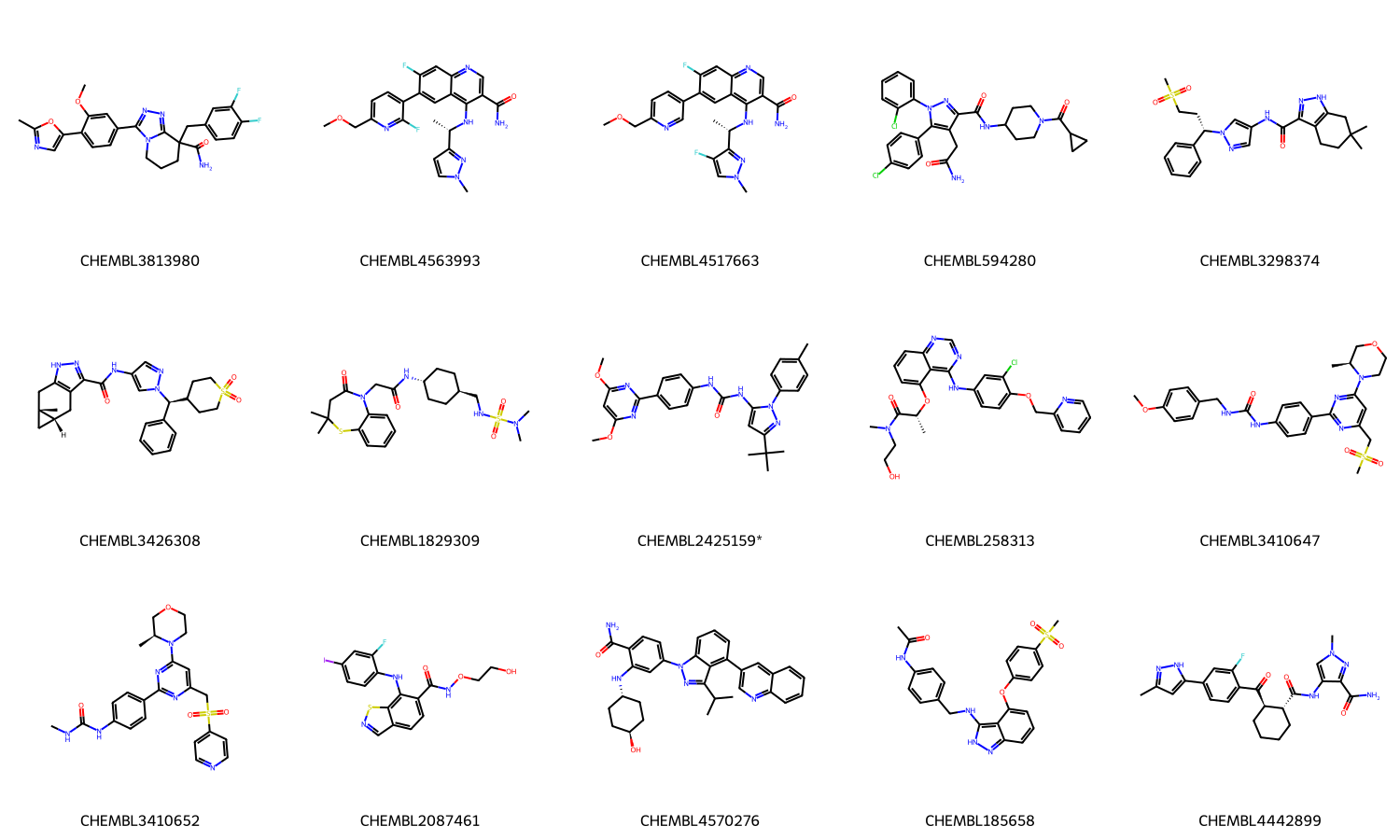}
    \caption{Structures of the CHEMBL compounds used for the logP calculations.}
    \label{SIfig:chembl_compounds}
\end{figure}
\clearpage

\section{Octanol Density}

\begin{figure}[H]
    \centering
    \includegraphics[width=0.5\linewidth]{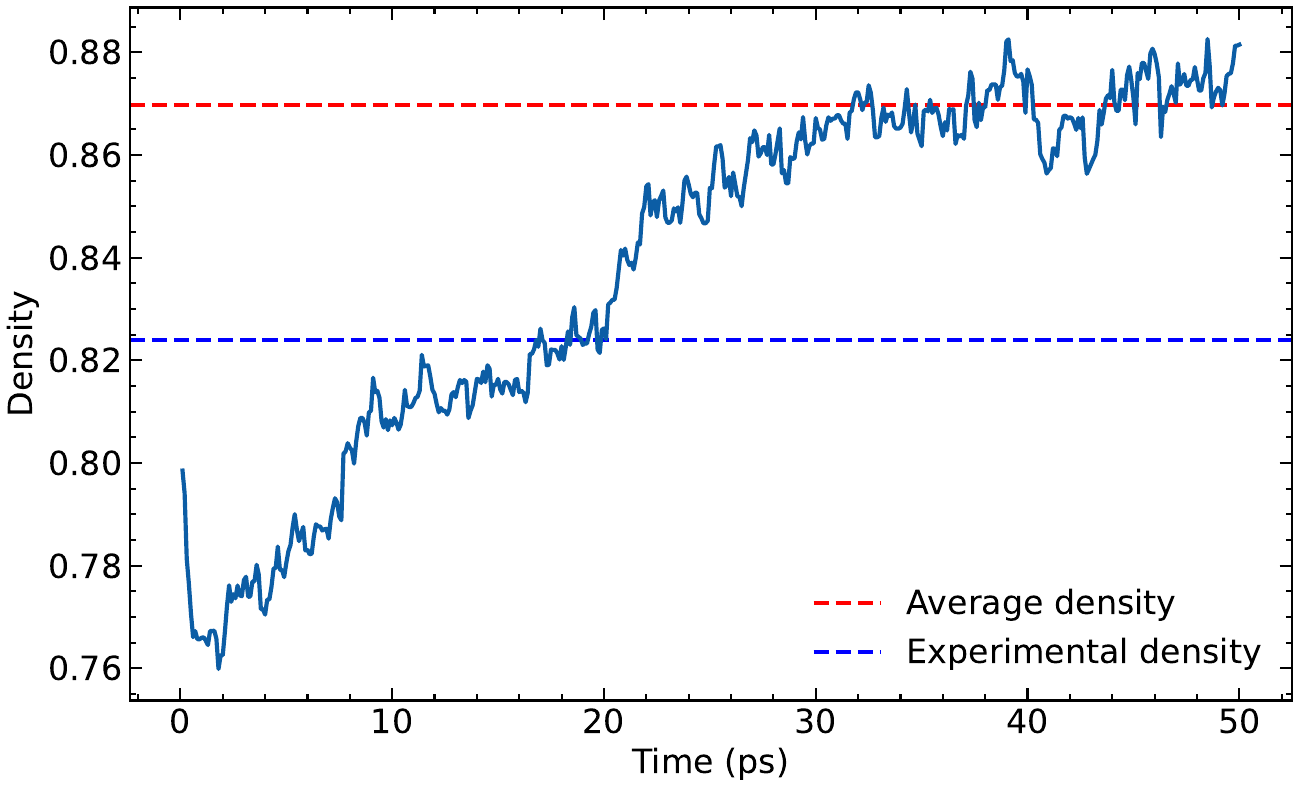}
    \caption{Density of liquid octanol simulated by \texttt{MACE-OFF24-SC}. Average density was calculated from the last 20 ps of the simulation}
    \label{SIfig:octanol_density}
\end{figure}
\clearpage

\section{Example GROMACS MDP file}
\label{SIsec:gmx_mdp}

\begin{lstlisting}[caption=mdp file for production GROMACS calculations for GAFF2/ABCG2 calculations]

title       = Absolute Solvation Free Energy in Octanol
integrator  = sd
nsteps      = 250000    ; 500 ps per window
dt          = 0.002
nstenergy   = 500
nstlog      = 500
nstxout-compressed = 500

continuation = yes
constraint_algorithm = lincs
constraints = h-bonds
lincs_iter  = 1
lincs_order = 4

nstlist         = 40
cutoff-scheme   = Verlet
ns_type         = grid
coulombtype     = PME
rcoulomb        = 1.2
rvdw            = 1.2
pbc             = xyz
disp_corr   = EnerPres

tc-grps     = System
tau_t       = 0.1
ref_t       = 298.15

pcoupl      = Parrinello-Rahman
pcoupltype  = isotropic
tau_p       = 2.0
ref_p       = 1.0
compressibility = 4.5e-5

; Free energy control
free_energy     = yes
init_lambda_state = 0   ; will be overwritten
delta_lambda    = 0
calc_lambda_neighbors = 1 ; calculate delta H to neighbors

; Lambda vectors
coul_lambdas = 0.00 0.14 0.29 0.43 0.57 0.71 0.86 1.00 1.00 1.00 1.00 1.00 1.00 1.00 1.00 1.00 1.00 1.00 1.00 1.00 1.00 1.00 1.00 1.00
vdw_lambdas  = 0.00 0.00 0.00 0.00 0.00 0.00 0.00 0.00 0.06 0.12 0.19 0.25 0.31 0.37 0.44 0.50 0.56 0.62 0.69 0.75 0.81 0.87 0.94 1.00

couple-moltype  = mol_009
couple-lambda0  = vdw-q
couple-lambda1  = none
couple-intramol = yes

sc-alpha    = 0.5
sc-power    = 1
sc-sigma    = 0.3
nstdhdl     = 100

\end{lstlisting}
\clearpage

